\documentclass[aps,prb,twocolumn,showpacs,a4paper]{revtex4}
\usepackage{amsmath,amssymb,amsfonts}
\usepackage{graphicx}

\newcommand{\ham}{\mathcal{H}}
\newcommand{\gas}{\mathsf{G}}
\newcommand{\open}{\mathsf{O}}
\newcommand{\closed}{\mathsf{C}}

\begin{document}

\title
{
  Thermodynamic anomalies in a lattice model of water
}
\author{M. Pretti and C. Buzano}
\affiliation
{
  Istituto Nazionale per la Fisica della Materia (INFM)
  and Dipartimento di Fisica, \\ Politecnico di Torino,
  Corso Duca degli Abruzzi 24, I-10129 Torino, Italy
}
\date{\today}
\begin{abstract}
We investigate a lattice-fluid model of water, defined on a
three-dimensional body centered cubic lattice. Model molecules
possess a tetrahedral symmetry, with four equivalent bonding arms,
aiming to mimic the formation of hydrogen bonds. The model is
similar to the one proposed by Roberts and Debenedetti [J. Chem.
Phys. {\bf 105}, 658 (1996)], simplified in that no distinction
between bond ``donors'' and ``acceptors'' is imposed. Bond
formation depends both on orientation and local density. In the
ground state, we show that two different ordered (ice) phases are
allowed. At finite temperature, we analyze homogeneous phases
only, working out phase diagram, response functions, the
temperature of maximum density locus, and the Kauzmann line. We
make use of a generalized first order approximation on a
tetrahedral cluster. In the liquid phase, the model exhibits
several anomalous properties observed in real water. In the low
temperature region (supercooled liquid), there are evidences of a
second critical point and, for some range of parameter values,
this scenario is compatible with the existence of a reentrant
spinodal.
\end{abstract}


\maketitle

\section{Introduction}

It is a well known fact that several thermodynamic properties of
water exhibit some anomalous
behavior~\cite{EisenbergKauzmann1969,Franks1982,Stanley2003}.
First of all, the heat capacity is unusually large and, at
ordinary pressures, the solid phase (ice) is less dense than the
liquid. Moreover, the liquid phase displays a temperature of
maximum density at constant pressure, while both isothermal
compressibility and isobaric heat capacity have a minimum as a
function of temperature. Generally speaking, the anomalous
properties can be explained by the ability of water molecules to
form hydrogen bonds, and by the peculiar features of such kind of
bonds~\cite{Stanley1998,Poole1994}. The same physics is thought to
underly the unusual properties of water as a solvent for apolar
compounds~\cite{FrankEvans1945,Stillinger1980}, that is of the
hydrophobic effect, whose importance in biophysics has been
recognized in the latest years~\cite{Dill1990}. Nevertheless, a
comprehensive theory which explains all of these phenomena has not
been developed yet.

``Realistic'' simulations of
water~\cite{StillingerRahman1974,Jorgensen1983,MahoneyJorgensen2000,Stanley2002},
based on more and more refined (but still phenomenological)
interaction potentials, have reached quite a high level of
accuracy in describing water thermodynamics. Nevertheless, they
are intrinsically limited by the large computational effort
required, which becomes still larger when it is necessary to
determine multiple derivatives of the free energy, such as
response functions. Moreover, due to the high level of microscopic
detail, both of geometry and of interactions, often they do not
make it easy to discriminate what is essential to explain
macroscopic properties. On the contrary, simplified models need
simpler numerical calculations and, even if their quantitative
accuracy is often poor, it is generally easier to trace
connections between microscopic interactions and macroscopic
properties~\cite{BellLavis1970,BenNaim1971,Bell1972,Lavis1973,BellSalt1976,%
LavisChristou1977,LavisChristou1979,%
MeijerKikuchiVanRoyen1982,%
HuckabyHanna1987,%
SastrySciortinoStanley1993jcp,%
RobertsDebenedetti1996,%
SilversteinHaymetDill1998}. A simplified mechanism, proposed to
account for the significant anomalies of water is the following
one (see for instance Refs.~\onlinecite{Stanley1994,Poole1994}).
The formation of a hydrogen bond requires that two molecules
assume certain relative orientations, staying at a distance larger
than the one needed to minimize Van der Waals energy. This fact
gives rise to a competition between the two kinds of interaction.
Optimizing Van der Waals interaction allows {\em higher density}
and {\em higher orientational entropy}, but yields a {\em weaker
binding energy}, whereas, optimizing hydrogen bonding requires a
{\em lower density} and a {\em lower orientational entropy}, but
gives rise to a {\em stronger binding energy}. Therefore, at low
enough temperature, local density and entropy fluctuations may
become positively correlated, thus rationalizing a change of sign
of the thermal expansion coefficient, that is a density maximum.
Such a simple mechanism has been implemented by different models,
both
on-~\cite{SastrySciortinoStanley1993jcp,%
RobertsDebenedetti1996,PatrykiejewPizioSokolowski1999,BruscoliniPelizzolaCasetti2002,%
BuzanoDestefanisPelizzolaPretti2003} and
off-lattice~\cite{SilversteinHaymetDill1998}, in
3~\cite{SastrySciortinoStanley1993jcp,%
RobertsDebenedetti1996} as well as 2
dimensions~\cite{SilversteinHaymetDill1998,PatrykiejewPizioSokolowski1999,BruscoliniPelizzolaCasetti2002,BuzanoDestefanisPelizzolaPretti2003}.

One of them is the 3-dimensional model proposed by Roberts and
Debenedetti
(RD)~\cite{RobertsDebenedetti1996,RobertsPanagiotopoulosDebenedetti1996},
defined on the body centered cubic lattice. Model molecules
possess four bonding arms (two donors and two acceptors) arranged
in a tetrahedral symmetry. Working on a lattice, one has to resort
to a trick to describe hydrogen bond weakening, when the two
participating molecules are too close to each other. Such a trick
is defined as follows. The energy of any formed bond is increased
of some fraction (weakened bond) by the presence of a third
molecule on a site close to the bond. Let us notice that the model
has the same bonding properties as the early model proposed by
Bell~\cite{Bell1972}, but the weakening criterion is different.
The RD model is quite appealing in that it has been shown to
predict some of real water thermodynamic anomalies, such as the
temperature of maximum density, also showing evidence of a
liquid-liquid phase separation in the supercooled region, and of a
second critical point. In view of investigations on mixtures of
water with other chemical species, as is the case, for instance,
in most biological processes, it would be desirable to obtain an
even simpler model, capable of capturing the same essential
features. As it has been pointed out by other
authors~\cite{SilversteinHaymetDill1998,SilversteinHaymetDill1999},
the distinction between donors and acceptors is likely to be not
so crucial to describe the physics of hydrogen bonding. Therefore,
a simplified version of the RD model, without a distinction
between hydrogen bond donors and acceptors, might be a good
compromise between simplicity and accuracy.

In this paper we investigate such a model, with a twofold purpose.
As mentioned above, we are meant to explore the possibility of
obtaining a simpler model with the same underlying physical
mechanism, and with qualitatively the same macroscopic properties.
In addition, we are interested in providing a more detailed
analysis of the effect of the weakening parameter, which turns out
to be extremely relevant to determine the phase diagram, mainly in
the supercooled liquid region. The paper is organized as follows.
In Sec.~II we define the model and analyze its ground state. In
Sec.~III we introduce the first order approximation in a cluster
variational formulation (cluster-site approximation), which we
employ for the analysis. Sec.~IV describes the results and Sec.~V
is devoted to some concluding remarks. An Appendix reports the
calculation of density response functions and spinodals for the
liquid phase.

\section{The model and the ground state}

\begin{figure}[t!]
  \includegraphics*[20mm,193mm][100mm,268mm]{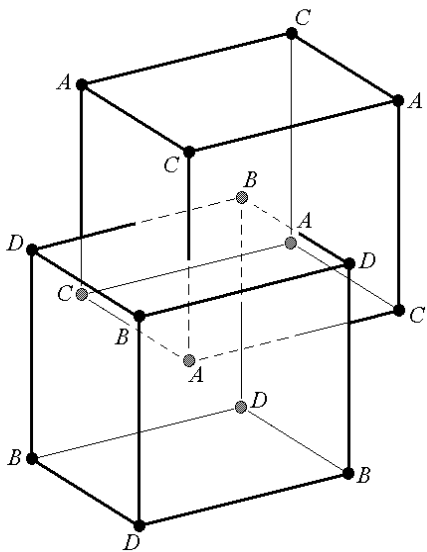}
  \caption
  {
    Two conventional cells of the body centered cubic lattice:
    $A,B,C,D$ denote 4 interpenetrating face centered cubic
    sublattices.
  }
  \label{fig:reticolo}
\end{figure}
\begin{figure}[t!]
  \includegraphics*[20mm,193mm][100mm,268mm]{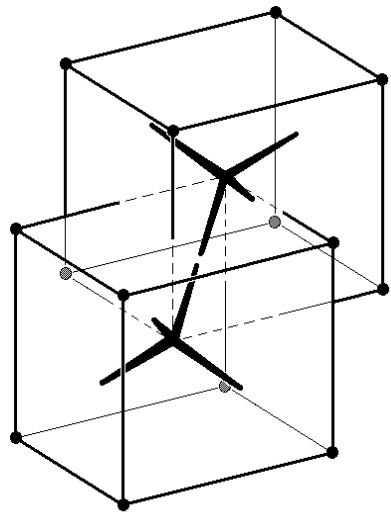}
  \caption
  {
    Two model molecules forming a H~bond.
    The lower molecule is in the $i=1$ configuration,
    the upper one is in the $i=2$ configuration.
  }
  \label{fig:molecole}
\end{figure}

Let us first introduce the model. Molecules are placed on the
sites of a body centered cubic lattice, whose structure is
sketched in Fig.~\ref{fig:reticolo}. A site may be empty or
occupied by a water molecule. An attractive potential energy
$-\epsilon<0$ is assigned to any pair of nearest neighbor (NN)
occupied sites. This is the ordinary Van der Waals contribution.
In the RD model, water molecules possess four arms that can form
hydrogen~(H) bonds (two donors and two acceptors), arranged in a
tetrahedral symmetry, so that they can point towards 4~out~of~8
NNs of a given site. We assume for simplicity that donors and
acceptors are undistinguishable, that is a H~bond is formed
whenever two NN molecules have a bonding arm pointing to each
other, yielding an energy~$-\eta<0$. Without such a distinction,
it turns out that a water molecule has only 2 different
configurations in which it can form H~bonds (see
Fig.~\ref{fig:molecole}). We assume that $w$~more configurations
are allowed, in which the molecule cannot form bonds. The
$w$~parameter is related to the bond-breaking entropy. Moreover,
to account for the fact that H bonds are most favorably formed
when water molecules are located at a certain distance, larger
than the optimal Van der Waals distance, the RD model assigns an
energy increase~$\eta c/6$, with $c\in[0,1]$, for each of the $6$
sites closest to the bond occupied by a water molecule (i.e., 3
out of 6 second neighbors of each participating molecule). A bond
surrounded by all $6$ water molecules is ``fully weakened'' and
contributes an energy~$-\eta (1-c)$.

\begin{figure}[t!]
  \resizebox{80mm}{!}{\includegraphics*[20mm,195mm][90mm,277mm]{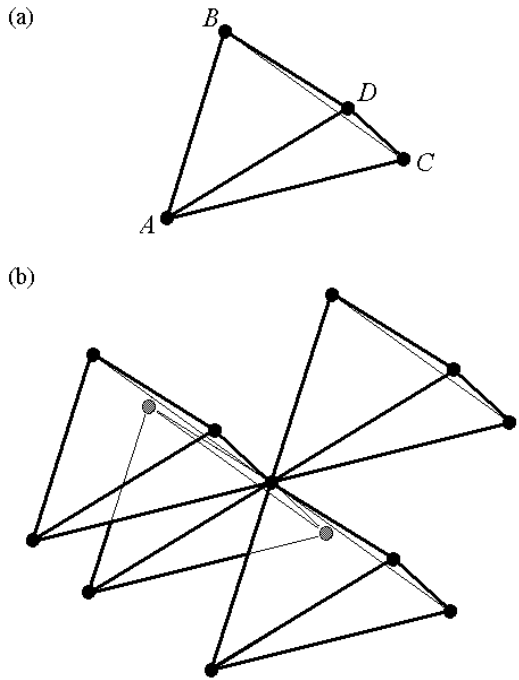}}
  \caption
  {
    (a) Basic cluster (irregular tetrahedron):
    $A,B,C,D$ denote sites in the 4 corresponding sublattices. $AB$,
    $BC$, $CD$, and $DA$ are NN pairs; $AC$ and $BD$ are second neighbor
    pairs.
    (b) Husimi tree structure corresponding to the generalized first
    order approximation on the tetrahedron.
  }
  \label{fig:cactustetraedro}
\end{figure}
The hamiltonian of the system can be written as a sum over
irregular tetrahedra, whose vertices lie on 4 different
face-centered cubic sublattices, shown in Fig.~\ref{fig:reticolo}.
One of such tetrahedra is shown in
Fig.~\ref{fig:cactustetraedro}(a). We have
\begin{equation}
  \ham =
  \frac{1}{6} \sum_{\langle \alpha,\beta,\gamma,\delta \rangle}
  \ham_{i_\alpha i_\beta i_\gamma i_\delta}
  ,
  \label{eq:ham}
\end{equation}
where $\ham_{ijkl}$ is a contribution which will be referred to as
tetrahedron hamiltonian, and the subscripts
$i_\alpha,i_\beta,i_\gamma,i_\delta$ label site configurations for
the 4 vertices $\alpha,\beta,\gamma,\delta$, respectively.
Possible configurations are: Empty site ($i=0$), site occupied by
a molecule in one of 2 bonding orientations ($i=1,2$; see
Fig.~\ref{fig:molecole}) or in one of $w$ non-bonding
configurations ($i=3$). Assuming that $(i,j)$, $(j,k)$, $(k,l)$,
and $(l,i)$ refer to NN pair configurations, the tetrahedron
hamiltonian reads
\begin{eqnarray}
  \ham_{ijkl} & = & -\epsilon(n_i n_j + n_j n_k + n_k n_l + n_l n_i)
  \\ &&
  -\eta \biggr[
  h_{ij} \left( 1 - c \frac{n_k + n_l}{2} \right) +
  h_{jk} \left( 1 - c \frac{n_l + n_i}{2} \right)
  \nonumber \\ && \ \ \ +
  h_{kl} \left( 1 - c \frac{n_i + n_j}{2} \right) +
  h_{li} \left( 1 - c \frac{n_j + n_k}{2} \right)
  \biggr]
  \nonumber
  ,
  \label{eq:tetraham}
\end{eqnarray}
where $n_i$ is an occupation variable, defined as $n_i=0$ for
$i=0$ (empty site) and $n_i=1$ otherwise (occupied site), while
$h_{ij}=1$ if the pair configuration $(i,j)$ forms a H~bond, and
$h_{ij}=0$ otherwise. Let us also assume that $i,j,k,l$ (in this
order) denote configurations of sites placed on, say, $A,B,C,D$
sublattices respectively. If $A,B,C,D$ are defined as in
Fig.~\ref{fig:reticolo}, we can define $h_{ij}=1$ if $i=1$ and
$j=2$, and $h_{ij}=0$ otherwise. With the above assumptions, the
tetrahedron hamiltonian is independent of the orientation, that is
of the arrangement of $A,B,C,D$ on its vertices. This is a nice
effect of removing the donor-acceptor distinction, which
considerably simplifies the whole analysis. Let us notice that
both Van der Waals ($-\epsilon n_i n_j$) and H~bond energies
($-\eta h_{ij}$), which are 2-body terms, are split among $6$
tetrahedra, whence the $1/6$~prefactor in Eq.~\eqref{eq:ham}. On
the contrary, the 3-body weakening terms ($\eta c h_{ij} n_k/6$)
are split between $2$ tetrahedra, thus the $1/6$ factor is
absorbed in the prefactor, while a $1/2$ factor is left in the
tetrahedron hamiltonian. Let us denote the tetrahedron
configuration probability by $p_{ijkl}$, with the same convention
about the subscript order, and assume that the probability
distribution is equal for every tetrahedron. Taking into account
that there are $6$ tetrahedra per site, we can write the following
expression for the internal energy per site of an infinite lattice
\begin{equation}
  u = \sum_{i=0}^3 w_i \sum_{j=0}^3 w_j \sum_{k=0}^3 w_k
  \sum_{l=0}^3 w_l p_{ijkl} \ham_{ijkl}.
  \label{eq:intenergy}
\end{equation}
The multiplicity for the tetrahedron configuration $(i,j,k,l)$ is
given by $w_i w_j w_k w_l$, where $w_i = w$ for $i=3$ (non-bonding
configuration) and $w_i = 1$ otherwise (bonding configuration or
vacancy).

Let us now have a look at the ground state of the model. In order
to do so, let us investigate the zero temperature grand-canonical
free energy $\omega^\circ = u - \mu \rho$ ($\mu$~being the
chemical potential and $\rho$ the density, i.e., the average site
occupation probability), which can be formally written in the same
way as the internal energy~$u$ of Eq.~\eqref{eq:intenergy},
replacing the tetrahedron hamiltonian~$\ham_{ijkl}$ by
\begin{equation}
  \tilde{\ham}_{ijkl} =
  \ham_{ijkl} - \mu \frac{n_i + n_j + n_k + n_l}{4}
  \label{eq:tetrahamtilde}
  .
\end{equation}
First of all, we have an infinitely dilute ``gas'' phase~($\gas$)
with zero density and zero free energy. Secondly, we can conceive
an ordered ``open'' ice phase~($\open$) with two fully occupied
sublattices, say $A$ and $B$, while the other sublattices, $C$ and
$D$, are empty ($\rho = 1/2$). In this configuration all molecules
are fully bonded, i.e., there are 2~bonds per molecule, and no
bond is weakened, so that the free energy turns out to be
\begin{equation}
  \omega^\circ_\open = -\epsilon -\eta - \mu/2
  .
  \label{eq:openicefreeenergy}
\end{equation}
Another possibility is the ``closed'' ice phase~($\closed$), in
which all sites are occupied ($\rho = 1$), the maximum number of
H~bonds is formed (for instance, all $AB$ and $CD$ pairs are
bonded), but all bonds are fully weakened. The resulting free
energy is
\begin{equation}
  \omega^\circ_\closed = -4\epsilon -2\eta(1 - c) -\mu
  .
  \label{eq:closedicefreeenergy}
\end{equation}
It is easy to show that the $\gas$~phase is thermodynamically
favored ($0<\omega^\circ_\open$ and $0<\omega^\circ_\closed$) for
$\mu < \mu_\mathrm{\gas\open}$, where
\begin{equation}
  \mu_{\gas \open} = -2\epsilon -2\eta
  ,
\end{equation}
the $\open$~phase is favored ($\omega^\circ_\open<0$ and
$\omega^\circ_\open<\omega^\circ_\closed$) for
$\mu_\mathrm{\gas\open} < \mu < \mu_{\open\closed}$, where
\begin{equation}
  \mu_{\open\closed} = -6\epsilon -2\eta(1-2c)
  ,
\end{equation}
and the $\closed$~phase is favored ($\omega^\circ_{\closed}<0$ and
$\omega^\circ_{\closed}<\omega^\circ_{\open}$) for $\mu >
\mu_{\open\closed}$. The $\open$~phase has actually a stability
region, i.e., $\mu_{\gas\open} < \mu_{\open\closed}$, provided
\begin{equation}
  c > \epsilon / \eta
  .
  \label{eq:ccondition}
\end{equation}
We shall always work in the latter regime, which allows to
reproduce two different forms of ice. Even if in the following we
shall not deal with ordered phases at finite temperature, the
latter choice should be the most reasonable one, in order to
describe real water properties. We have considered also the
possibility of different structured phases, respectively with 1 or
3 occupied sublattices, but they never turn out to be stable, in
the physical range of parameter values.

\section{First order approximation}

We perform the finite temperature analysis by means of a
generalized first order approximation on a tetrahedron cluster.
Let us introduce the approximation in the framework of the cluster
variation method, an improved mean-field theory which in principle
can take into account correlations at arbitrarily large, though
finite, distances. In Kikuchi's original work~\cite{Kikuchi1951},
an approximate entropy expression was obtained by heuristic
arguments, while, in more recent and rigorous
formulations~\cite{An1988}, the approximation is shown to be
equivalent to a truncation of a cluster cumulant expansion of the
entropy. The approximation is expected to work, because of a rapid
decreasing of the cumulant magnitude, upon increasing the cluster
size, namely when the latter becomes larger than the correlation
length of the system~\cite{Morita1972}. A particular approximation
is defined by the largest clusters left in the truncated
expansion, usually denoted as basic clusters. One obtains a free
energy functional in the cluster probability distributions, to be
minimized, according to the variational principle of statistical
mechanics.

For our model we choose a number of irregular tetrahedra as basic
clusters, namely 4 out of 24 tetrahedra sharing a given site, as
sketched in Fig.~\ref{fig:cactustetraedro}. This choice actually
turns out to coincide with the (generalized) first order
approximation (on the tetrahedron cluster), which is also
equivalent to an exact calculation on a Husimi
lattice~\cite{Pretti2003}, whose (tetrahedral) building blocks are
just arranged as in Fig.~\ref{fig:cactustetraedro}(b). Such an
approximation has not only the advantage of high simplicity, due
to the fact that the only clusters retained in the expansion are
basic clusters and single sites (it is sometimes referred to as
cluster-site approximation~\cite{Oates1999}), but also of a
relative accuracy, which has been recognized for different models,
even with orientation dependent
interactions~\cite{BuzanoDestefanisPelizzolaPretti2003}. Let us
notice that the internal energy is treated exactly, because the
range of interactions does not exceed the basic cluster size. The
grand canonical free energy per site $\omega = u - Ts - \mu \rho$
($T$ being the temperature and $s$ the entropy per site) can be
written as
\begin{eqnarray}
  \frac{\omega}{T} & = &
  \sum_{i=0}^3 w_i \sum_{j=0}^3 w_j \sum_{k=0}^3 w_k \sum_{l=0}^3 w_l p_{ijkl}
  \left( \frac{\tilde{\ham}_{ijkl}}{T} + \ln p_{ijkl} \right)
  \nonumber \\ &&
  - 3 \sum_{i=0}^3 w_i p_i \ln p_i
  ,
  \label{eq:func}
\end{eqnarray}
where $p_i$ is the probability of the site configuration~$i$
(temperature is expressed in energy units, whence entropy in
natural units). In this paper we focus on liquid, i.e.,
homogeneous state, hence we assume that all sites have the same
configuration probability distribution. The latter can then be
obtained as a marginal of the tetrahedron distribution~$p_{ijkl}$,
by the following symmetrized expression
\begin{equation}
  p_i =
  \sum_{j=0}^3 w_j \sum_{k=0}^3 w_k \sum_{l=0}^3 w_l
  \frac{p_{ijkl} + p_{lijk} + p_{klij} + p_{jkli}}{4}
  .
  \label{eq:marginal}
\end{equation}
The free energy turns out to be a function of the only tetrahedron
probability distribution, taken as variational parameter. The
minimization with respect to such parameter, with the
normalization constraint
\begin{equation}
  \sum_{i=0}^3 w_i \sum_{j=0}^3 w_j \sum_{k=0}^3 w_k \sum_{l=0}^3 w_l
  p_{ijkl} = 1
  ,
  \label{eq:constraint}
\end{equation}
can be performed by the Lagrange multiplier method, yielding the
equations
\begin{equation}
  p_{ijkl} = \xi^{-1} e^{-\tilde{\ham}_{ijkl}/T}
  \left( p_i p_j p_k p_l \right)^{3/4}
  ,
  \label{eq:cvmeq}
\end{equation}
where $\xi$, related to the Lagrange multiplier, can be computed
by imposing the constraint Eq.~\eqref{eq:constraint} as
\begin{equation}
  \xi =
  \sum_{i=0}^3 w_i \sum_{j=0}^3 w_j \sum_{k=0}^3 w_k \sum_{l=0}^3 w_l
  e^{-\tilde{\ham}_{ijkl}/T}
  \left( p_i p_j p_k p_l \right)^{3/4}
  .
  \label{eq:costnorm}
\end{equation}
Eq.~\eqref{eq:cvmeq} is in a fixed point form, and can be solved
numerically by simple iteration (natural iteration
method~\cite{Kikuchi1974}). For the cluster-site approximation,
the numerical procedure can be proved to reduce the free energy at
each iteration~\cite{Kikuchi1974,Pretti2003}, and therefore to
converge to local minima. Let us notice that the symmetrized
marginalization Eq.~\eqref{eq:marginal} imposes implicitly a
homogeneity constraint, useful to investigate the metastable
liquid properties. In fact it is easy to see that
Eq.~\eqref{eq:cvmeq}, due to invariance of $\tilde{\ham}_{ijkl}$
under cycle permutation of the subscripts [see
Eqs.~\eqref{eq:tetraham} and~\eqref{eq:tetrahamtilde}], determines
a tetrahedron distribution $p_{ijkl}$ with the same property,
giving rise to four equal site marginals. Therefore, we must be
aware that the stationary point we find may be unstable with
respect to a translational symmetry breaking. Anyway, from the
solution of Eq.~\eqref{eq:cvmeq} one obtains a tetrahedron
probability distribution $\{p_{ijkl}\}$, whence one can compute
the thermal average of every observable, the internal energy by
Eq.~\eqref{eq:intenergy} and the free energy by
Eq.~\eqref{eq:func}. The latter can be also related to the
normalization constant as
\begin{equation}
  \omega = - T \ln \xi
  ,
  \label{eq:enlibcostnorm}
\end{equation}
whence $\xi$ can be viewed as an effective (single site) grand
canonical partition function. We shall make use of this property
in the following.

\section{Thermodynamic properties}

In order to investigate the model properties, let us fix a set of
parameters. First of all, let us take $\epsilon/\eta = 0.25$. This
value is equal to the one employed for a previous mean field
analysis~\cite{BuzanoPretti2003jcp}, and similar to the one chosen
by RD for the original model~\cite{RobertsDebenedetti1996}. This
choice accounts for the greater binding energy of hydrogen bonds
with respect to Van der Waals interactions, and will be kept fixed
throughout the present analysis. As far as the multiplicity of
non-bonding water configurations is concerned, we set $w = 20$, to
mimic the high directionality of hydrogen bonds. From the phase
diagram analysis, it turns out that it is necessary to set this
parameter large enough to let anomalous properties appear, but
further increase does not change qualitatively the phase behavior.
Therefore, also the latter parameter will be held fixed in the
following. On the contrary, we shall investigate in detail the
effect of changing the weakening parameter $c$, which is actually
the crucial one for the present model, in the regime $c>0.25$,
according to Eq.~\eqref{eq:ccondition}.

\subsection{Phase diagrams}

\begin{figure}[t!]
  \includegraphics*[40mm,85mm][120mm,250mm]{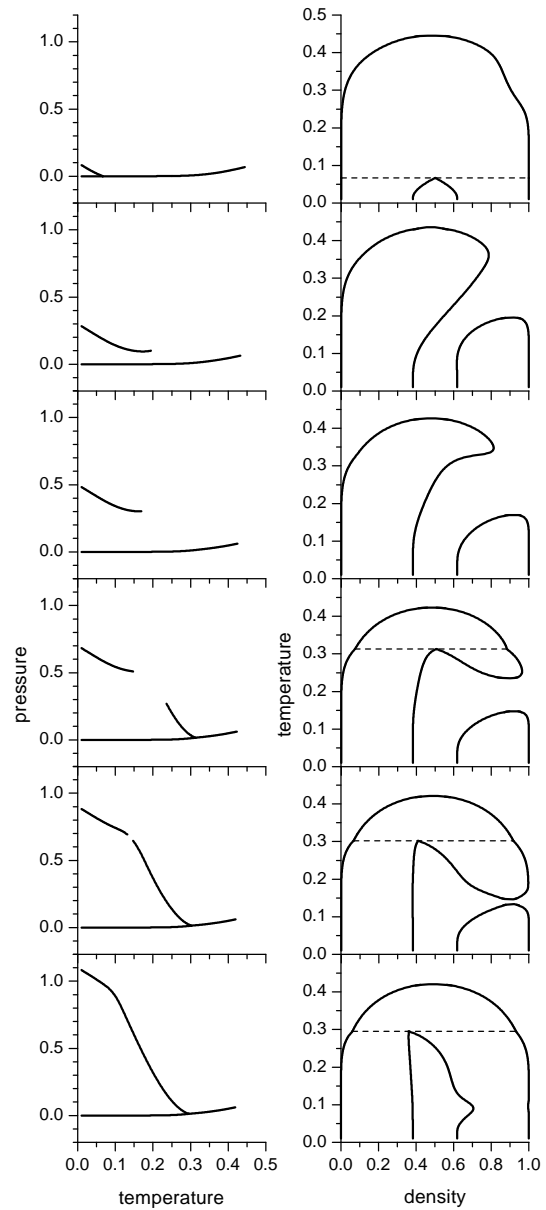}
  \caption
  {
    Pressure ($P/\eta$) vs temperature ($T/\eta$) phase diagrams (left column) and
    temperature vs density ($\rho$) phase diagrams (right column)
    for $\epsilon/\eta=0.25$ and $w=20$. From top to bottom
    $c=0.3,0.4,0.5,0.6,0.7,0.8$, respectively.
    Thick solid lines denote (first order) phase transitions
    (left) and delimit coexistence regions (right).
    A thin dashed line (right) corresponds to three-phase coexistence (triple point).
  }
  \label{fig:e025w20}
\end{figure}
Let us report in Fig.~\ref{fig:e025w20} a sequence of
temperature-pressure (left column) and density-temperature (right
column) phase diagrams, for different $c$ values. Imposing
homogeneity, our analysis includes both thermodynamically stable
and metastable (supercooled) phases, even if stability is not
investigated. Let us notice that pressure can be determined as $P
= -\omega$, due to the fact that the free energy has been defined
as a grand-canonical potential. We have assumed volume per site
equal to~1, i.e., pressure is expressed in energy units.

For $c=0.3$ we have essentially an ordinary liquid-vapor
coexistence line, terminating at a critical point, even if an
anomalous temperature dependence of the liquid density (without a
maximum) can be observed. In the very low temperature region, an
intermediate density liquid phase appear, giving rise to a triple
point. This phase is actually an unphysical solution of our
equations, in that the entropy turns out to be negative. In this
region a crystalline phase is likely to be stable.

For $c=0.4$ the phase diagram undergoes a dramatic change. The
intermediate liquid phase region becomes a unique region with the
ordinary liquid, and the triple point is replaced by a second
critical point, where the low-high density liquid coexistence
terminates. The second critical point still lies in a negative
entropy region. The density of the liquid coexisting with the
vapor displays a maximum as a function of temperature.

For $c=0.5$ the density maximum becomes more pronounced, but there
is no topological change in the phase diagram.

For $c=0.6$ a new coexistence line appears between the high
density and the low density liquid. Such a transition line
(probably metastable) is negatively sloped in the
temperature-pressure phase diagram, and terminates at a third
critical point. In this scenario, the ordinary liquid phase
stability is delimited by a reentrant spinodal, as will be
verified in the following. A similar scenario has been already
observed in a two-dimensional
model~\cite{BuzanoDestefanisPelizzolaPretti2003}.

For $c=0.7$ the two ``metastable'' critical points gets closer,
and finally, for $c=0.8$, they merge, giving rise to a unique
high-low density liquid coexistence, which terminates at a triple
point. The topology of the phase diagram is again similar to the
one obtained for low $c$ values, but in this case there exists a
region where the low density liquid has positive entropy. From a
microscopical point of view, such a liquid phase is characterized
by a high probability of bonding configurations, i.e., it is a
highly hydrogen bonded phase. This feature corresponds to a lower
density, as previously pointed out.

The role of the weakening parameter is well characterized by the
sequence of temperature-pressure phase diagrams. Actually, the
general trend is that, upon increasing $c$, the stability region
of the low density liquid (having a low number of weakening
molecules) gets larger and larger. In the next part of this work
we focus on two particular parameter choices ($c=0.4,0.6$), as
representatives of a range of $c$ values in which the model
results are qualitatively consistent with the anomalies observed
for real liquid water in ordinary temperature and pressure
conditions. The general analysis reported above is completed by
studying the locus of density extrema (maxima and minima) as a
function of temperature, the liquid phase spinodals, and the
Kauzmann line, where the liquid entropy vanishes (ideal glass
transition).
\begin{figure}[t!]
  \includegraphics*[35mm,127mm][115mm,250mm]{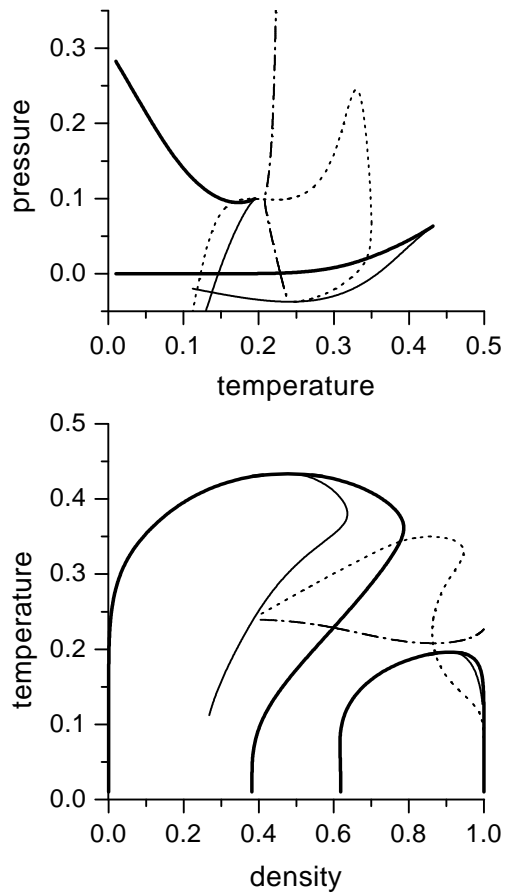}
  \caption
  {
    Pressure ($P/\eta$) vs temperature ($T/\eta$) phase diagram (top panel) and
    temperature vs density ($\rho$) phase diagram (bottom panel)
    for $\epsilon/\eta=0.25$, $w=20$, $c=0.4$.
    Thick solid lines denote (first order) phase transitions
    (top) and delimit coexistence regions (bottom).
    Thin (solid, dotted, and dash-dotted) lines denote
    spinodals, TED locus, and Kauzmann line, respectively.
  }
  \label{fig:e025w20c040}
\end{figure}
\begin{figure}[t!]
  \includegraphics*[35mm,87mm][115mm,250mm]{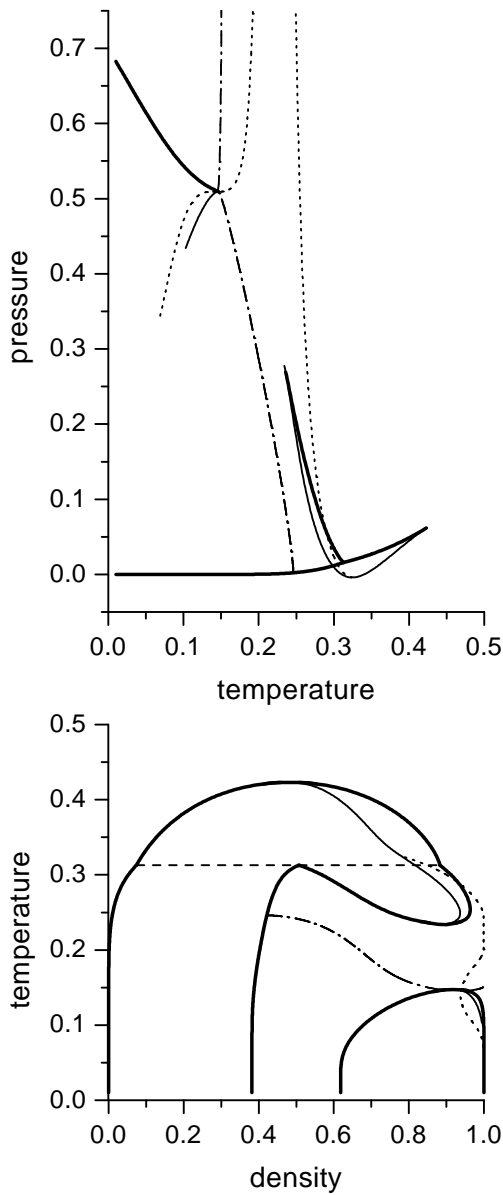}
  \caption
  {
    Pressure ($P/\eta$) vs temperature ($T/\eta$) phase diagram (top panel) and
    temperature vs density ($\rho$) phase diagram (bottom panel)
    for $\epsilon/\eta=0.25$, $w=20$, $c=0.6$.
    Thick solid lines denote (first order) phase transitions
    (top) and delimit coexistence regions (bottom).
    A thin dashed line (bottom) corresponds to three-phase coexistence (triple point).
    Thin (solid, dotted, and dash-dotted) lines denote
    spinodals, TED locus, and Kauzmann line, respectively.
  }
  \label{fig:e025w20c060}
\end{figure}

\subsection{TED locus, spinodal, and Kauzmann line}

One of the thermodynamic anomalies of the present model is the
temperature of maximum density (TMD) along isobars for the liquid
phase. Nevertheless, for some pressure range, it is possible to
observe also a temperature of minimum density, therefore we shall
generally speak about a temperature of extremum density (TED).
Joining temperatures of maximum (or extremum) density at different
pressures defines the so called TMD (or TED) locus. At ordinary
pressure, the TMD locus is a negatively sloped line in the $T$-$P$
phase diagram of real water. In principle, we could determine the
TED~locus numerically, by adjusting the chemical potential in
order to fix the pressure and then imposing that the (isobaric)
thermal expansion coefficient vanishes. Actually we have performed
a different calculation, based on the effective partition
function~\eqref{eq:enlibcostnorm}, rewritten as a function of only
two variational parameters, namely, the density $\rho$ and the
fraction $\phi$ of bonding molecules. Details about this
calculation, which allows to determine density response functions
and spinodals as well, are given in the Appendix. The limit of
stability of the liquid phase (spinodal) is the locus in which the
metastable liquid ceases to be a minimum of the free energy, and
becomes a saddle point. Actually, in the homogeneity hypothesis,
we add a constraint to the free energy, thus neglecting stability
loss with respect to symmetry broken (ice) phases. Aware of this,
spinodals can be obtained by imposing that the hessian determinant
of the homogeneous free energy (or partition function) vanishes,
as shown in the Appendix. Let us notice that, in this case, it is
not possible to work out the result making use of the natural
iteration Eqs.~\eqref{eq:cvmeq}, because the attraction basin of
the liquid phase gets smaller and smaller, and vanishes at the
spinodal. On the contrary, the locus at which the liquid phase
entropy vanishes (Kauzmann line), can be easily determined
numerically.

\begin{figure}[t!]
  \includegraphics*[35mm,196mm][115mm,250mm]{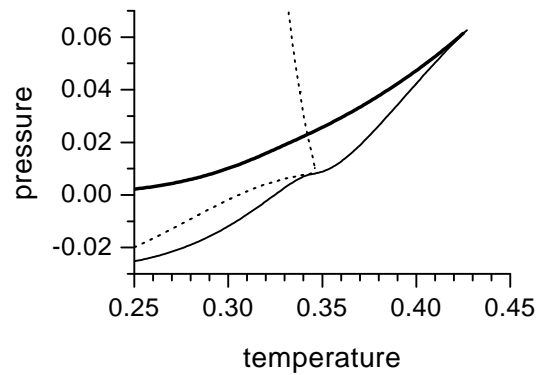}
  \caption
  {
    Pressure ($P/\eta$) vs temperature ($T/\eta$) phase diagram
    for $\epsilon/\eta=0.25$, $w=20$, $c=0.5$.
    A thick solid lines denotes the liquid-vapor transition.
    The thin (solid and dotted) lines denote
    the liquid spinodal and the TMD locus, respectively.
  }
  \label{fig:e025w20c050op}
\end{figure}

The results are shown in Figs.~\ref{fig:e025w20c040}
and~\ref{fig:e025w20c060} for $c=0.4$ and $c=0.6$, respectively.
In both cases, we find a density maximum as a function of
temperature for liquid coexisting with vapor (and at constant
pressure as well), and the TMD slightly decreases upon increasing
pressure. Another common feature is the presence of a high-low
density liquid coexistence, and of a corresponding critical point.
For $c=0.4$ the critical point lies in the negative entropy
region, beneath the Kauzmann line, while for $c=0.6$ it lies in
the positive entropy region. The Kauzmann line displays a cusp
(towards low temperature) in the vicinity of the critical point.
The low approximation level of our analysis does not allow to take
all of this information as reliable, but let us notice that a
similar scenario has been predicted also by a statistical analysis
of the potential energy landscape of simulated water, performed by
Sciortino and coworkers on the basis of the inherent structure
theory and of some simplifying
assumptions~\cite{SciortinoLaNaveTartaglia2003}. As far as the TMD
locus is concerned, we observe a peculiar maximum in the $T$-$P$
plane. For $c=0.6$ the maximum occurs at a far higher pressure
than for $c=0.4$. After the maximum, upon decreasing temperature,
the TMD locus becomes a temperature of minimum density locus,
ending in the second critical point. After that, the line
continues (again as a TMD locus) in the metastability region of
the high density liquid with respect to the low density liquid.
Upon decreasing pressure, we find the most relevant differences
between the two cases. For $c=0.4$, the TMD reaches a maximum and
then decreases again, as we move in the negative pressure region.
For $c=0.6$, the TMD always increases. In both cases the TMD locus
terminates against the liquid-vapor spinodal. In the $T$-$P$
diagram, the meeting point corresponds to a minimum of the
spinodal line, as required by the thermodynamic consistency
arguments of Speedy and
Debenedetti~\cite{Speedy1982I,Speedy1982II,Speedy1987,DebenedettiDantonio1986I,DantonioDebenedetti1987}.
Such a minimum is placed at a much higher temperature for $c=0.6$
than for $c=0.4$. As far as the liquid-vapor spinodal is
concerned, another important difference is observed, namely, for
$c=0.6$ the spinodal reenters the positive pressure region, ending
in the above mentioned ``third'' critical point. Such a reentrance
gives rise to a divergence in the response functions, measured at
constant pressure, as we shall see below. This is not the case for
$c=0.4$. As one could expect, it is possible to verify that the
spinodal reenters the positive pressure region, if and only if the
third critical point exists. The boundary between the two regimes
is found to be around (actually a little greater than) $c=0.5$. A
region of the $T$-$P$ phase diagram for $c=0.5$, showing the
reentrance of the TMD locus, is reported in
Fig.~\ref{fig:e025w20c050op}.

\subsection{Response functions}

\begin{figure}[t!]
  \includegraphics*[35mm,116mm][115mm,250mm]{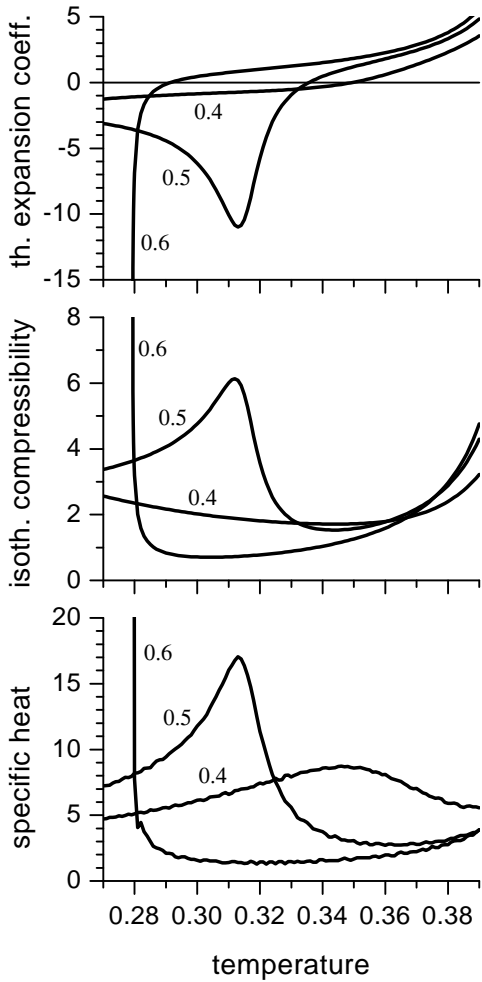}
  \caption
  {
    Response functions at constant pressure ($P/\eta=0.05$)
    for the liquid phase as a function of temperature ($T/\eta$),
    for $\epsilon/\eta=0.25$ and $w=20$.
    From top to bottom we show
    the isobaric thermal expansion coefficient ($\eta \alpha_P$),
    the isothermal compressibility ($\eta \kappa_T$),
    and the specific heat ($c_P$).
    Numerals beside each plot denote $c$ values.
  }
  \label{fig:pcost}
\end{figure}
Let us now investigate the density response functions and the
specific heat of the liquid at constant pressure $P/\eta=0.05$,
roughly corresponding to $1/10$ of the liquid-vapor critical
pressure. Also for this calculations, details are left to the
Appendix. We find anomalous behavior, similar to that of real
liquid water. The first response function we consider is the
thermal expansion coefficient $\alpha_P =
(-\partial\ln\rho/\partial T)_P$, which, from statistical
mechanics, is known to be proportional to the entropy-volume
cross-correlation. For ordinary fluids, $\alpha_P$~is always
positive, i.e., the local entropy and the local specific volume
are positively correlated. On the contrary, for our model
$\alpha_P$ (Fig.~\ref{fig:pcost}, top panel) is anomalous. As
temperature is lowered, the expansion coefficient vanishes (at the
TMD), and then becomes negative. For $c=0.4$, we have observed
just a shallow minimum (not shown) around $T/\eta=0.17$. The
minimum is more pronounced for $c=0.5$, which, as mentioned above,
is still in a regime where the third critical point does not
exists. Finally, for $c=0.6$, the minimum becomes a divergence,
due to the spinodal reentrance in the positive pressure region. Of
course, the divergent behavior can be observed only for pressure
values less than the third critical point pressure. The trend of
the isothermal compressibility $\kappa_T =
(\partial\ln\rho/\partial P)_T$ is also anomalous
(Fig.~\ref{fig:pcost}, middle panel). For a typical liquid,
$\kappa_T$ decreases as one lowers temperature, because it is
proportional to density fluctuations, whose magnitude decreases,
upon decreasing temperature. On the contrary, we can observe that
$\kappa_T$, once reached a minimum, begins to increase upon
decreasing temperature. Only a broad maximum is observed for
$c=0.4$; the maximum becomes sharper for $c=0.5$, and finally
becomes a divergence for $c=0.6$. The constant pressure specific
heat $c_P = (-T\partial^2 \mu/\partial T^2)_P$
(Fig.~\ref{fig:pcost}, bottom panel) displays a completely
analogous behavior, with the minimum occurring at a higher
temperature. Qualitatively similar thermodynamic anomalies are
observed in real liquid water, even if the possibility of
divergent-like behavior in the supercooled regime is no longer
believed to be realistic~\cite{Stanley2003}.

\subsection{A comparison with the RD model}

\begin{figure}[t!]
  \includegraphics*[35mm,191mm][115mm,250mm]{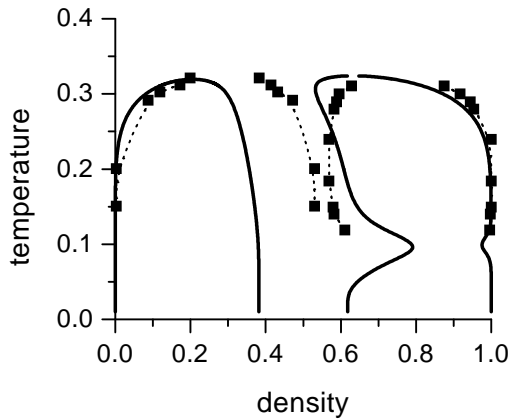}
  \caption
  {
    Temperature ($T/\eta$) vs. density ($\rho$) phase diagram
    for $\epsilon/\eta=0.2$, $w=16$, $c=0.8$ (solid lines),
    compared to Monte Carlo results from
    Ref.~\protect\onlinecite{RobertsPanagiotopoulosDebenedetti1996}
    (scatters). Dotted lines are eyeguides.
  }
  \label{fig:e020w16c080}
\end{figure}

Let us finally report the results of a comparison with the RD
model, of which, as previously mentioned, the present model is a
simplified version. In particular, let us consider the results of
a Monte Carlo
simulation~\cite{RobertsPanagiotopoulosDebenedetti1996}, in which
parameters were chosen to push the low-high density liquid
critical point to a temperature of the same order of magnitude of
the ordinary liquid-vapor critical point. In this work, evidence
was given that the model actually predicts liquid-liquid
coexistence, and that the latter is not an artifact of
approximations. The model parameters are $\epsilon/\eta=0.2$,
$c=0.8$, while the total number of molecule orientations is
$q=108$. In order to a comparison, the latter parameter is to be
renormalized, in that RD model molecules, possessing bonding arms
of two different kinds (donors and acceptors), allow $12$
different bonding configurations, while the present model allows
only $2$. We have performed the renormalization as follows
\begin{equation}
  \frac{q}{12} = \frac{w+2}{2}
  ,
\end{equation}
obtaining $w=16$. The results are reported in
Fig.~\ref{fig:e020w16c080}. The topology of the
(temperature-density) phase diagram is qualitatively similar, and
also the quantitative agreement of critical temperatures is good.
Only the density of the liquid coexisting with the vapor displays
a significant discrepancy. The reentrance in the liquid-liquid
coexistence region at low temperature is reproduced, even if RD
conjecture that a lower critical point exists, mainly on the basis
of first order approximation
results~\cite{RobertsDebenedetti1996}, whereas we find coexistence
down to zero temperature. Anyway, as shown before, the presence or
absence of a lower critical point may be induced by small $c$
value variations.

\section{Discussion and conclusions}

In this paper we have investigated a lattice fluid model with
water-like features, by a generalized first order approximation.
As mentioned in the Introduction, the mechanism by which our model
describes water anomalies is essentially based on the competition
between the Van der Waals isotropic interaction and the highly
directional H~bonding interaction, and on the difference between
the respective optimal interaction distances. In the lattice
framework, the latter is taken into account by means of the
weakening trick, proposed by RD~\cite{RobertsDebenedetti1996}. We
have actually simplified the original RD model by neglecting the
distinction between H~bond donors and acceptors. In spite of this,
the model turns to reproduce qualitatively several thermodynamic
anomalies of real water at constant pressure: a maximum of
density, and a minimum of isothermal compressibility and specific
heat. This fact confirms the idea that the distinction between
donors and acceptors is not crucial to the physics of
H~bonding~\cite{SilversteinHaymetDill1998}. By a suitable
parameter scaling, we have also compared some results of the
simplified model with those of the original model obtained by
Monte Carlo simulations, and we have verified that the
simplification and the approximation seem to preserve the
qualitative features of the phase diagram, though in a case of
quite simple topology. On the simplified model, it is possible to
perform a generalized first order approximation on a tetrahedron
cluster, without introducing further ``ad hoc'' approximations, as
it was necessary in the original paper by
RD~\cite{RobertsDebenedetti1996}. Moreover, with respect to Monte
Carlo simulations, the first order approximation has not only the
advantage of a much lower computational effort, but also allows a
well defined extrapolation of the equation of state in the
supercooled regime, without the need of criteria to prevent the
simulation dynamics from falling into crystalline states. In fact,
analysis of the model predictions in the metastable phase region
is extremely interesting to rationalize observed thermodynamic
anomalies.

As previously shown, there is evidence of a second (metastable)
critical point, which terminates a line of first order transitions
between two liquid phases at different densities. For $c \lesssim
0.5$, the second critical point lies at a temperature lower than
the Kauzmann temperature, at which the configurational entropy
vanishes. Moreover, the Kauzmann line displays a reentrance in the
vicinity of the second critical point. All of these features turn
out to be in a remarkably good agreement with the simplified
statistical analysis of the potential energy landscape of
simulated water, recently performed by Sciortino and
coworkers~\cite{SciortinoLaNaveTartaglia2003}.

We have also computed the temperature of extremum density locus
and the liquid spinodals, in the framework of our approximation.
In the ordinary temperature and pressure region, the TED locus is
a negatively sloped line (in the $T$-$P$ diagram) corresponding to
a density maximum, as observed in experiments. The locus displays
a pressure maximum, and, for higher pressure values, the liquid
becomes normal. After the maximum, upon decreasing temperature,
the TED line denotes a minimum density, which is a peculiar
feature of our model. The TED line crosses the Kauzmann line very
close to its reentrance, as recently predicted by
Speedy~\cite{Speedy2002}. On the other side, the TED line is
reentrant for low $c$ values and it is not for higher $c$ values.
In both cases it terminates at a minimum of the spinodal in the
$T$-$P$ plane, as required by thermodynamic
consistency~\cite{Speedy1982I}. Nevertheless, in the former case
the spinodal does not reenter the positive pressure half-plane,
while in the latter case it does. In this case, response functions
exhibit a divergent behavior. Such results are interesting in that
the reentrant spinodal was one of the conjectures invoked to
explain thermodynamic anomalies of
water~\cite{Speedy1982I,ZhengDurbenWolfAngell1991}. The fact, that
the same model with different parameter values may predict or not
a reentrant spinodal, suggests that in real water there is
probably a subtle balance of interactions, which does not allow to
discriminate easily between the two possible scenarios. This fact
has been previously pointed out by phenomenological
models~\cite{Poole1994}, but, to our knowledge, not yet by
microscopical models. The most recent and accurate molecular
dynamics simulations suggest a scenario with a non-reentrant
spinodal and a reentrant (``nose shaped'') TMD
locus~\cite{Stanley2003}. The present model accounts for such a
scenario for $c \approx 0.5$. In this case, a slight reentrance of
the spinodal, induced by the vicinity of the TMD locus, can be
observed. A similar ``intermediate'' situation is consistent with
the results of an analytical equation of state derived by Truskett
et al.~\cite{TruskettDebenedettiSastryTorquato1999}. We have
remarked this feature in that it may justify the early
experimental results which supported the reentrant spinodal
conjecture~\cite{ZhengDurbenWolfAngell1991}.

As a conclusion, let us remind that the model in the present
treatment is not able to provide microscopic structural details as
simulations do, but its most appealing feature is simplicity. In
spite of how little we put in the model, the latter turns out to
give a qualitatively correct description of the peculiar
thermodynamics of water, and is consistent with predictions based
on much more sophisticated models and simulations. Therefore, we
suggest that it may be suitable as a starting point to investigate
complex phenomena, such as the solvation of apolar solutes, in
which the physics of hydrogen bonding plays a significant role.

\appendix*
\section{Spinodals and response functions}

In this appendix we report in detail the calculation of the
spinodal lines and of the density response functions, i.e., the
(isobaric) thermal expansion coefficient $\alpha_P$ and the
isothermal compressibility $\kappa_T$. The (isobaric) heat
capacity $c_P$ is determined by a numerical derivative of the
chemical potential $\mu$, which comes out in a natural way from
the same calculation, as a function of temperature and pressure.

Eq.~\eqref{eq:enlibcostnorm} suggests that the normalization
constant~$\xi$ [Eq.~\eqref{eq:costnorm}] can be used as a
variational form of the effective (single site) grand canonical
partition function, to be maximized with respect to the site
probability distribution~$p_i$. We can also observe that the
homogeneous liquid phase is isotropic, whence the 2 bonding
configurations on each site must have the same probability
\begin{equation}
  p_1 = p_2
  .
\end{equation}
With the above assumption, we can replace the fourfold summation
over the configurations $i,j,k,l$ of a tetrahedron with a double
summation over the number $n=0,\dots,4$ of occupied sites and the
number $k=0,\dots,n$ of molecules in a bonding configuration.
There are ${4 \choose n} {n \choose k}$ ways of arranging
molecules with the above requirements, and the multiplicity of the
resulting configuration ($k$ bonding molecules and $n-k$ non
bonding ones) is $2^k w^{n-k}$, whence we obtain
\begin{equation}
  \xi = \sum_{n=0}^{4} \sum_{k=0}^{n} {4 \choose n} {n \choose k}
  2^k w^{n-k} \left( p_0^{4-n} p_1^k p_3^{n-k} \right)^{3/4}
  x_{n,k}
  ,
\end{equation}
where $x_{n,k}$ are linear combinations of the Boltzmann weights
$e^{-\tilde{\ham}_{ijkl}/T}$, appearing in Eq.~\eqref{eq:cvmeq}.
They are obtained as described in Tab.~\ref{tab:boltzmann}.
\begin{table}[t]
  \begin{ruledtabular}
  \begin{tabular}{cc|cc|l}
    $n$ & $k$ & ${4 \choose n}$ & ${n \choose k}$ & $x_{n,k}/z^{n/4}$ \\
    \hline
    \hline
    0 & 0 & 1 & 1 & 1 \\
    \hline
    1 & 0 & 4 & 1 & 1 \\
    1 & 1 & 4 & 1 & 1 \\
    \hline
    2 & 0 & 6 & 1 & $\frac{1}{3} + \frac{2}{3}e^{\epsilon/T}$ \\
    2 & 1 & 6 & 2 & $\frac{1}{3} + \frac{2}{3}e^{\epsilon/T}$ \\
    2 & 2 & 6 & 1 & $\frac{1}{3} + \frac{2}{3}e^{\epsilon/T} \left( \frac{3}{4} + \frac{1}{4}e^{\eta/T} \right)$ \\
    \hline
    3 & 0 & 4 & 1 & $e^{2 \epsilon/T}$ \\
    3 & 1 & 4 & 3 & $e^{2 \epsilon/T}$ \\
    3 & 2 & 4 & 3 & $e^{2 \epsilon/T} \left\{ \frac{1}{3} + \frac{2}{3} \left[ \frac{3}{4} + \frac{1}{4}e^{\eta (1-c/2)/T} \right] \right\}$ \\
    3 & 3 & 4 & 1 & $e^{2 \epsilon/T} \left[ \frac{1}{2} + \frac{1}{2}e^{\eta (1-c/2)/T} \right]$ \\
    \hline
    4 & 0 & 1 & 1 & $e^{4 \epsilon/T}$ \\
    4 & 1 & 1 & 4 & $e^{4 \epsilon/T}$ \\
    4 & 2 & 1 & 6 & $e^{4 \epsilon/T} \left\{ \frac{1}{3} + \frac{2}{3} \left[ \frac{3}{4} + \frac{1}{4}e^{\eta (1-c)/T} \right] \right\}$ \\
    4 & 3 & 1 & 4 & $e^{4 \epsilon/T} \left[ \frac{1}{2} + \frac{1}{2}e^{\eta (1-c)/T} \right]$ \\
    4 & 4 & 1 & 1 & $e^{4 \epsilon/T} \left[ \frac{1}{8} + \frac{6}{8}e^{\eta (1-c)/T} + \frac{1}{8}e^{2 \eta (1-c)/T} \right]$ \\
  \end{tabular}
  \end{ruledtabular}
  \caption{
    $x_{n,k}$ coefficients.
    The way they have been computed is explained hereafter.
    The chemical potential is taken into account by the
    $z^{n/4}$~prefactor, where $z \equiv e^{\mu/T}$ is the fugacity.
    For $n=0,1$ no interaction is possible,
    whence only the fugacity term is present.
    For $n=2$ there are 6 possible arrangements of molecules on
    the sites; in 2 of them ($1/3$) the 2 molecules are not NNs
    and there is no interaction; in the remaining 4 cases ($2/3$)
    the 2 molecules are NNs and interact with the Van der Waals
    energy (whence the factor $e^{\epsilon/T}$). Only when $k=2$
    a H bond can occur (whence the factor $e^{\eta/T}$),
    in 1 out of 4 possible configurations of 2
    bonding molecules, i.e., when the molecules point an arm to
    each other. The H bond is not weakened because the neighbor
    sites are empty.
    For $n=3$ all possible arrangements have 2 pairs of NN
    molecules, whence the factor $e^{2\epsilon/T}$.
    If $k=0,1$ no other interaction exists.
    If $k=2$ we can place the 2 bonding molecules on NN sites
    in $2/3$ of the cases, and in $1/4$ of their allowed
    configuration they form a H bond, which is half weakened,
    because one of the NN sites is occupied.
    With $k=3$ bonding molecules, there are $2^3=8$ possible
    configurations, 4 of which have 1 formed bond.
    For $n=4$ all sites are occupied and there are 4 NN pairs,
    whence the factor $e^{4\epsilon/T}$. With $k=0,1$ bonding
    molecules, no other interaction is possible. With $k=2$,
    there are $4/6=2/3$ arrangements in which the bonding molecules
    are placed on NN sites and form a H bond in $1/4$ of their
    configurations. The bond is fully weakened, because both the neighbor
    sites are occupied. With $k=3$ bonding molecules, the explanation
    is equivalent to the $n=3$ case. Finally, with $k=4$ bonding
    molecules, it is necessary to enumerate all the $2^4=16$
    configurations, 2 of which have no bond, 6 have 1 bond, and 2
    have 2 bonds.
  }
  \label{tab:boltzmann}
\end{table}
In order to avoid imposing the normalization constraint, it is
convenient to rewrite the site probabilities as a function of the
density $\rho$ (occupation probability) and the fraction $\phi$ of
molecules in a bonding configuration. It is easy to obtain
\begin{eqnarray}
  p_0 & = & 1 - \rho  \\
  p_1 & = & \frac{\rho \phi}{2} \\
  p_3 & = & \frac{\rho (1-\phi)}{w}
  ,
\end{eqnarray}
whence
\begin{equation}
  \xi =
  \sum_{n=0}^{4} {4 \choose n} f_{4,n}^{3/4}(\rho)
  \sum_{k=0}^{n} {n \choose k} \left( 2^k w^{n-k} \right)^{1/4} f_{n,k}^{3/4}(\phi)
  x_{n,k},
  \label{eq:partfun}
\end{equation}
where
\begin{equation}
  f_{n,k}(x) \equiv x^k (1-x)^{n-k}
  .
  \label{eq:f}
\end{equation}
In order to impose thermodynamic equilibrium, we have to minimize
$\xi$ with respect to the variational parameters $\rho$ and
$\phi$. Let us notice that $\xi$ also depends on the temperature
$T$ and the chemical potential $\mu$ (or equivalently the fugacity
$z=e^{\mu/T}$), as appropriate in the grand-canonical ensemble.
Such a dependence is hidden in the $x_{n,k}$ coefficients (see
Tab.~\ref{tab:boltzmann}). We impose the necessary minimum
condition, by setting the derivatives of $\xi$ with respect to
$\rho$ and $\phi$ to zero. Moreover, we are interested in working
at fixed pressure $P$, therefore we have to impose a third
equation $\omega=-P$, where, from Eq.~\eqref{eq:enlibcostnorm}, we
have $\omega=-T\log\xi$. We have thus to solve a system of three
equations
\begin{eqnarray}
  \frac{\partial \xi}{\partial (\ln \rho)} & = & 0
  \label{eq:xr0} \\
  \frac{\partial \xi}{\partial (\ln \phi)} & = & 0
  \label{eq:xf0} \\
  \xi - e^{P/T} & = & 0
  \label{eq:omegap}
  ,
\end{eqnarray}
which is actually an implicit definition of the three functions
$\rho(T,P)$, $\phi(T,P)$, and $z(T,P)$. Having determined
numerically the fugacity $z(T,P)$, we immediately obtain the
chemical potential $\mu = T \log z$, whence the specific heat $c_P
= (-T\partial^2 \mu/\partial T^2)_P$ by a numerical derivative.

Let us now consider the three simultaneous Eqs.~\eqref{eq:xr0},
\eqref{eq:xf0}, and~\eqref{eq:omegap}. As previously mentioned,
the left hand sides are functions of $\rho,\phi,z,T,P$. If we
replace $\rho,\phi,z$ by the previously determined functions of
$T,P$, we obtain three functions that are identically equal to
$0$, for each value of $T$ and $P$. Therefore, taking the partial
derivatives with respect to $X=T,P$, we obtain the following
linear system:
\begin{equation}
  \mathsf{A} \cdot
  \left( \begin{matrix}
    \displaystyle  \frac{\partial (\ln \rho)}{\partial X}
  \cr \cr
    \displaystyle  \frac{\partial (\ln \phi)}{\partial X}
  \cr \cr
    \displaystyle  \frac{\partial (\ln z)   }{\partial X}
  \end{matrix} \right)
  = -
  \left( \begin{matrix}
    \displaystyle
    \frac{\partial^2 \xi}{\partial (\ln \rho) \partial X}
  \cr \cr
    \displaystyle
    \frac{\partial^2 \xi}{\partial (\ln \phi) \partial X}
  \cr \cr
    \displaystyle
    \frac{\partial \xi}{\partial X}
    - e^{P/T} \frac{\partial (P/T)}{\partial X}
  \end{matrix} \right)
  ,
  \label{eq:hessiansystem}
\end{equation}
where
\begin{equation}
  \mathsf{A} =
  \left( \begin{matrix} \displaystyle
    \frac{\partial^2 \xi}{\partial (\ln \rho)^2}
    & \displaystyle
    \frac{\partial^2 \xi}{\partial (\ln \rho) \partial (\ln \phi)}
    & \displaystyle
    \frac{\partial^2 \xi}{\partial (\ln \rho) \partial (\ln z)}
  \cr & & \cr \displaystyle
    \frac{\partial^2 \xi}{\partial (\ln \phi) \partial (\ln \rho)}
    & \displaystyle
    \frac{\partial^2 \xi}{\partial (\ln \phi)^2}
    & \displaystyle
    \frac{\partial^2 \xi}{\partial (\ln \phi) \partial (\ln z)}
  \cr & & \cr \displaystyle
    \frac{\partial \xi}{\partial (\ln \rho)}
    & \displaystyle
    \frac{\partial \xi}{\partial (\ln \phi)}
    & \displaystyle
    \frac{\partial \xi}{\partial (\ln z)}
  \end{matrix} \right)
\end{equation}
and
\begin{eqnarray}
  \frac{\partial (P/T)}{\partial T}
  & = & - \frac{P}{T^2} \\
  & & \nonumber \\
  \frac{\partial (P/T)}{\partial P}
  & = & \frac{1}{T}
  .
\end{eqnarray}
The derivatives of $\xi$ can be easily determined from
Eq.~\eqref{eq:partfun}, making use of
\begin{eqnarray}
  \frac{d f_{n,k}^a(x)}{d (\ln x)}
  & = &
  \frac{a (k-nx)}{1-x} f_{n,k}^a(x)
  \\
  \frac{d^2 f_{n,k}^a(x)}{d (\ln x)^2}
  & = &
  \frac{a^2 (k-nx)^2 - a (n-k) x}{(1-x)^2} f_{n,k}^a(x)
  ,
  \nonumber \\ &&
\end{eqnarray}
obtained from Eq.~\eqref{eq:f}, $a$ being a generic exponent, and
of
\begin{equation}
  \frac{\partial x_{n,k}}{\partial (\ln z)} = \frac{n}{4} \, x_{n,k}
  ,
\end{equation}
while $\partial x_{n,k}/\partial T$ can be obtained from
Table~\ref{tab:boltzmann}. In order to determine the density
response function, we solve Eq.~\eqref{eq:hessiansystem} with
respect to $\partial (\ln \rho)/\partial X$, making use of the
Kramer's rule, yielding
\begin{equation}
  \frac{\partial (\ln \rho)}{\partial X} =
  \frac{\det \mathsf{B}}{\det \mathsf{A}}
  ,
\end{equation}
where $\mathsf{B}$ is obtained by replacing the right hand side of
Eq.~\eqref{eq:hessiansystem} in the first column of $\mathsf{A}$.
In this way the isobaric thermal expansion coefficient $\alpha_P =
-\partial (\ln \rho)/\partial T$ and the isothermal
compressibility $\kappa_T = \partial (\ln \rho)/\partial P$ are
expressed as a function of $\rho,\phi,z,T,P$. As previously
mentioned, $\rho,\phi,z$ can be determined numerically as a
function of $T,P$, for instance by solving the simultaneous
Eqs.~\eqref{eq:xr0},~\eqref{eq:xf0}, and~\eqref{eq:omegap}. As far
as the TED locus is concerned, we have to add a fourth equation,
imposing $\alpha_P=0$. Finally, spinodal lines can be obtained as
the locus in which the above response functions diverge, that is,
$\det\mathsf{A}=0$. Let us observe that, in the latter case, only
three simultaneous equations are needed. In fact, if the grand
canonical equilibrium conditions~\eqref{eq:xr0} and~\eqref{eq:xf0}
are verified, the first two elements of the third line of
$\mathsf{A}$ (which we may denote by $A_{31},A_{32}$) vanish,
therefore we can write $\det\mathsf{A} = A_{33}
\det\mathsf{A}_{33}$, where $\mathsf{A}_{33}$ is the submatrix of
$\mathsf{A}$ obtained by removing the third row and the third
column. As a consequence, the equation $\det\mathsf{A}=0$ does not
contain $P$. Solving the latter simultaneously with
Eqs.~\eqref{eq:xr0} and~\eqref{eq:xf0}, which do not depend on $P$
as well, allows to determine $\rho(T)$, $\phi(T)$, and $z(T)$
defining the spinodal line. Then one can determine $\xi(T)$ by
means of Eq.~\eqref{eq:partfun}, and finally $P(T)=T\ln\xi(T)$.


\begin{thebibliography}{46}
\expandafter\ifx\csname natexlab\endcsname\relax\def\natexlab#1{#1}\fi
\expandafter\ifx\csname bibnamefont\endcsname\relax
  \def\bibnamefont#1{#1}\fi
\expandafter\ifx\csname bibfnamefont\endcsname\relax
  \def\bibfnamefont#1{#1}\fi
\expandafter\ifx\csname citenamefont\endcsname\relax
  \def\citenamefont#1{#1}\fi
\expandafter\ifx\csname url\endcsname\relax
  \def\url#1{\texttt{#1}}\fi
\expandafter\ifx\csname urlprefix\endcsname\relax\def\urlprefix{URL }\fi
\providecommand{\bibinfo}[2]{#2}
\providecommand{\eprint}[2][]{\url{#2}}

\bibitem[{\citenamefont{Eisenberg and Kauzmann}(1969)}]{EisenbergKauzmann1969}
\bibinfo{author}{\bibfnamefont{D.}~\bibnamefont{Eisenberg}} \bibnamefont{and}
  \bibinfo{author}{\bibfnamefont{W.}~\bibnamefont{Kauzmann}},
  \emph{\bibinfo{title}{The Structure and Properties of Water}}
  (\bibinfo{publisher}{Oxford University Press, Oxford}, \bibinfo{year}{1969}).

\bibitem[{\citenamefont{Franks}(1982)}]{Franks1982}
\bibinfo{editor}{\bibfnamefont{F.}~\bibnamefont{Franks}}, ed.,
  \emph{\bibinfo{title}{Water: a Comprehensive Treatise}}
  (\bibinfo{publisher}{Plenum Press, New York}, \bibinfo{year}{1982}).

\bibitem[{\citenamefont{Stanley et~al.}(2003)}]{Stanley2003}
\bibinfo{author}{\bibfnamefont{H.~E.} \bibnamefont{Stanley}}
  \bibnamefont{et~al.}, \bibinfo{journal}{J. Stat. Phys.}
  \textbf{\bibinfo{volume}{110}}, \bibinfo{pages}{1039} (\bibinfo{year}{2003}).

\bibitem[{\citenamefont{Stanley et~al.}(1998)}]{Stanley1998}
\bibinfo{author}{\bibfnamefont{H.~E.} \bibnamefont{Stanley}}
  \bibnamefont{et~al.}, \bibinfo{journal}{Physica A}
  \textbf{\bibinfo{volume}{257}}, \bibinfo{pages}{213} (\bibinfo{year}{1998}).

\bibitem[{\citenamefont{Poole et~al.}(1994)\citenamefont{Poole, Sciortino,
  Grande, Stanley, and Angell}}]{Poole1994}
\bibinfo{author}{\bibfnamefont{P.~H.} \bibnamefont{Poole}},
  \bibinfo{author}{\bibfnamefont{F.}~\bibnamefont{Sciortino}},
  \bibinfo{author}{\bibfnamefont{T.}~\bibnamefont{Grande}},
  \bibinfo{author}{\bibfnamefont{H.~E.} \bibnamefont{Stanley}},
  \bibnamefont{and} \bibinfo{author}{\bibfnamefont{C.~A.}
  \bibnamefont{Angell}}, \bibinfo{journal}{Phys. Rev. Lett.}
  \textbf{\bibinfo{volume}{73}}, \bibinfo{pages}{1632} (\bibinfo{year}{1994}).

\bibitem[{\citenamefont{Frank and Evans}(1945)}]{FrankEvans1945}
\bibinfo{author}{\bibfnamefont{H.~S.} \bibnamefont{Frank}} \bibnamefont{and}
  \bibinfo{author}{\bibfnamefont{M.~W.} \bibnamefont{Evans}},
  \bibinfo{journal}{J. Chem. Phys.} \textbf{\bibinfo{volume}{13}},
  \bibinfo{pages}{507} (\bibinfo{year}{1945}).

\bibitem[{\citenamefont{Stillinger}(1980)}]{Stillinger1980}
\bibinfo{author}{\bibfnamefont{F.~H.} \bibnamefont{Stillinger}},
  \bibinfo{journal}{Science} \textbf{\bibinfo{volume}{209}},
  \bibinfo{pages}{451} (\bibinfo{year}{1980}).

\bibitem[{\citenamefont{Dill}(1990)}]{Dill1990}
\bibinfo{author}{\bibfnamefont{K.~A.} \bibnamefont{Dill}},
  \bibinfo{journal}{Biochemistry} \textbf{\bibinfo{volume}{29}},
  \bibinfo{pages}{7133} (\bibinfo{year}{1990}).

\bibitem[{\citenamefont{Stillinger and Rahman}(1974)}]{StillingerRahman1974}
\bibinfo{author}{\bibfnamefont{F.~H.} \bibnamefont{Stillinger}}
  \bibnamefont{and} \bibinfo{author}{\bibfnamefont{A.}~\bibnamefont{Rahman}},
  \bibinfo{journal}{J. Chem. Phys.} \textbf{\bibinfo{volume}{60}},
  \bibinfo{pages}{1545} (\bibinfo{year}{1974}).

\bibitem[{\citenamefont{Jorgensen et~al.}(1983)}]{Jorgensen1983}
\bibinfo{author}{\bibfnamefont{W.~L.} \bibnamefont{Jorgensen}}
  \bibnamefont{et~al.}, \bibinfo{journal}{J. Chem. Phys.}
  \textbf{\bibinfo{volume}{79}}, \bibinfo{pages}{926} (\bibinfo{year}{1983}).

\bibitem[{\citenamefont{Mahoney and Jorgensen}(2000)}]{MahoneyJorgensen2000}
\bibinfo{author}{\bibfnamefont{M.~W.} \bibnamefont{Mahoney}} \bibnamefont{and}
  \bibinfo{author}{\bibfnamefont{W.~L.} \bibnamefont{Jorgensen}},
  \bibinfo{journal}{J. Chem. Phys.} \textbf{\bibinfo{volume}{112}},
  \bibinfo{pages}{8910} (\bibinfo{year}{2000}).

\bibitem[{\citenamefont{Stanley et~al.}(2002)}]{Stanley2002}
\bibinfo{author}{\bibfnamefont{H.~E.} \bibnamefont{Stanley}}
  \bibnamefont{et~al.}, \bibinfo{journal}{Physica A}
  \textbf{\bibinfo{volume}{315}}, \bibinfo{pages}{281} (\bibinfo{year}{2002}).

\bibitem[{\citenamefont{Bell and Lavis}(1970)}]{BellLavis1970}
\bibinfo{author}{\bibfnamefont{G.~M.} \bibnamefont{Bell}} \bibnamefont{and}
  \bibinfo{author}{\bibfnamefont{D.~A.} \bibnamefont{Lavis}},
  \bibinfo{journal}{J. Phys. A} \textbf{\bibinfo{volume}{3}},
  \bibinfo{pages}{568} (\bibinfo{year}{1970}).

\bibitem[{\citenamefont{Ben-Naim}(1971)}]{BenNaim1971}
\bibinfo{author}{\bibfnamefont{A.}~\bibnamefont{Ben-Naim}},
  \bibinfo{journal}{J. Chem. Phys.} \textbf{\bibinfo{volume}{54}},
  \bibinfo{pages}{3682} (\bibinfo{year}{1971}).

\bibitem[{\citenamefont{Bell}(1972)}]{Bell1972}
\bibinfo{author}{\bibfnamefont{G.~M.} \bibnamefont{Bell}}, \bibinfo{journal}{J.
  Phys. C} \textbf{\bibinfo{volume}{5}}, \bibinfo{pages}{889}
  (\bibinfo{year}{1972}).

\bibitem[{\citenamefont{Lavis}(1973)}]{Lavis1973}
\bibinfo{author}{\bibfnamefont{D.~A.} \bibnamefont{Lavis}},
  \bibinfo{journal}{J. Phys. C} \textbf{\bibinfo{volume}{6}},
  \bibinfo{pages}{1530} (\bibinfo{year}{1973}).

\bibitem[{\citenamefont{Bell and Salt}(1976)}]{BellSalt1976}
\bibinfo{author}{\bibfnamefont{G.~M.} \bibnamefont{Bell}} \bibnamefont{and}
  \bibinfo{author}{\bibfnamefont{D.~W.} \bibnamefont{Salt}},
  \bibinfo{journal}{J. Chem. Soc., Faraday Trans. II}
  \textbf{\bibinfo{volume}{72}}, \bibinfo{pages}{76} (\bibinfo{year}{1976}).

\bibitem[{\citenamefont{Lavis and Christou}(1977)}]{LavisChristou1977}
\bibinfo{author}{\bibfnamefont{D.~A.} \bibnamefont{Lavis}} \bibnamefont{and}
  \bibinfo{author}{\bibfnamefont{N.~I.} \bibnamefont{Christou}},
  \bibinfo{journal}{J. Phys. A} \textbf{\bibinfo{volume}{10}},
  \bibinfo{pages}{2153} (\bibinfo{year}{1977}).

\bibitem[{\citenamefont{Lavis and Christou}(1979)}]{LavisChristou1979}
\bibinfo{author}{\bibfnamefont{D.~A.} \bibnamefont{Lavis}} \bibnamefont{and}
  \bibinfo{author}{\bibfnamefont{N.~I.} \bibnamefont{Christou}},
  \bibinfo{journal}{J. Phys. A} \textbf{\bibinfo{volume}{12}},
  \bibinfo{pages}{1869} (\bibinfo{year}{1979}).

\bibitem[{\citenamefont{Meijer et~al.}(1982)\citenamefont{Meijer, Kikuchi, and
  Royen}}]{MeijerKikuchiVanRoyen1982}
\bibinfo{author}{\bibfnamefont{P.~H.~E.} \bibnamefont{Meijer}},
  \bibinfo{author}{\bibfnamefont{R.}~\bibnamefont{Kikuchi}}, \bibnamefont{and}
  \bibinfo{author}{\bibfnamefont{E.~V.} \bibnamefont{Royen}},
  \bibinfo{journal}{Physica A} \textbf{\bibinfo{volume}{115}},
  \bibinfo{pages}{124} (\bibinfo{year}{1982}).

\bibitem[{\citenamefont{Huckaby and Hanna}(1987)}]{HuckabyHanna1987}
\bibinfo{author}{\bibfnamefont{D.~A.} \bibnamefont{Huckaby}} \bibnamefont{and}
  \bibinfo{author}{\bibfnamefont{R.~S.} \bibnamefont{Hanna}},
  \bibinfo{journal}{J. Phys. A} \textbf{\bibinfo{volume}{20}},
  \bibinfo{pages}{5311} (\bibinfo{year}{1987}).

\bibitem[{\citenamefont{Sastry et~al.}(1993)\citenamefont{Sastry, Sciortino,
  and Stanley}}]{SastrySciortinoStanley1993jcp}
\bibinfo{author}{\bibfnamefont{S.}~\bibnamefont{Sastry}},
  \bibinfo{author}{\bibfnamefont{F.}~\bibnamefont{Sciortino}},
  \bibnamefont{and} \bibinfo{author}{\bibfnamefont{H.~E.}
  \bibnamefont{Stanley}}, \bibinfo{journal}{J. Chem. Phys.}
  \textbf{\bibinfo{volume}{98}}, \bibinfo{pages}{9863} (\bibinfo{year}{1993}).

\bibitem[{\citenamefont{Roberts and
  Debenedetti}(1996)}]{RobertsDebenedetti1996}
\bibinfo{author}{\bibfnamefont{C.~J.} \bibnamefont{Roberts}} \bibnamefont{and}
  \bibinfo{author}{\bibfnamefont{P.~G.} \bibnamefont{Debenedetti}},
  \bibinfo{journal}{J. Chem. Phys.} \textbf{\bibinfo{volume}{105}},
  \bibinfo{pages}{658} (\bibinfo{year}{1996}).

\bibitem[{\citenamefont{Silverstein et~al.}(1998)\citenamefont{Silverstein,
  Haymet, and Dill}}]{SilversteinHaymetDill1998}
\bibinfo{author}{\bibfnamefont{K.~A.~T.} \bibnamefont{Silverstein}},
  \bibinfo{author}{\bibfnamefont{A.~D.~J.} \bibnamefont{Haymet}},
  \bibnamefont{and} \bibinfo{author}{\bibfnamefont{K.~A.} \bibnamefont{Dill}},
  \bibinfo{journal}{J. Am. Chem. Soc.} \textbf{\bibinfo{volume}{120}},
  \bibinfo{pages}{3166} (\bibinfo{year}{1998}).

\bibitem[{\citenamefont{Stanley et~al.}(1994)}]{Stanley1994}
\bibinfo{author}{\bibfnamefont{H.~E.} \bibnamefont{Stanley}}
  \bibnamefont{et~al.}, \bibinfo{journal}{Physica A}
  \textbf{\bibinfo{volume}{205}}, \bibinfo{pages}{122} (\bibinfo{year}{1994}).

\bibitem[{\citenamefont{Patrykiejew et~al.}(1999)\citenamefont{Patrykiejew,
  Pizio, and Soko{\l}owski}}]{PatrykiejewPizioSokolowski1999}
\bibinfo{author}{\bibfnamefont{A.}~\bibnamefont{Patrykiejew}},
  \bibinfo{author}{\bibfnamefont{O.}~\bibnamefont{Pizio}}, \bibnamefont{and}
  \bibinfo{author}{\bibfnamefont{S.}~\bibnamefont{Soko{\l}owski}},
  \bibinfo{journal}{Phys. Rev. Lett.} \textbf{\bibinfo{volume}{83}},
  \bibinfo{pages}{3442} (\bibinfo{year}{1999}).

\bibitem[{\citenamefont{Bruscolini et~al.}(2002)\citenamefont{Bruscolini,
  Pelizzola, and Casetti}}]{BruscoliniPelizzolaCasetti2002}
\bibinfo{author}{\bibfnamefont{P.}~\bibnamefont{Bruscolini}},
  \bibinfo{author}{\bibfnamefont{A.}~\bibnamefont{Pelizzola}},
  \bibnamefont{and} \bibinfo{author}{\bibfnamefont{L.}~\bibnamefont{Casetti}},
  \bibinfo{journal}{Phys. Rev. Lett.} \textbf{\bibinfo{volume}{88}},
  \bibinfo{pages}{089601} (\bibinfo{year}{2002}).

\bibitem[{\citenamefont{Buzano et~al.}(2003)\citenamefont{Buzano, de~Stefanis,
  Pelizzola, and Pretti}}]{BuzanoDestefanisPelizzolaPretti2003}
\bibinfo{author}{\bibfnamefont{C.}~\bibnamefont{Buzano}},
  \bibinfo{author}{\bibfnamefont{E.}~\bibnamefont{de~Stefanis}},
  \bibinfo{author}{\bibfnamefont{A.}~\bibnamefont{Pelizzola}},
  \bibnamefont{and} \bibinfo{author}{\bibfnamefont{M.}~\bibnamefont{Pretti}},
  \bibinfo{journal}{Phys. Rev. E} \textbf{\bibinfo{volume}{69}},
  \bibinfo{pages}{(to be published)} (\bibinfo{year}{2003}).

\bibitem[{\citenamefont{Roberts et~al.}(1996)\citenamefont{Roberts,
  Panagiotopoulos, and Debenedetti}}]{RobertsPanagiotopoulosDebenedetti1996}
\bibinfo{author}{\bibfnamefont{C.~J.} \bibnamefont{Roberts}},
  \bibinfo{author}{\bibfnamefont{A.~Z.} \bibnamefont{Panagiotopoulos}},
  \bibnamefont{and} \bibinfo{author}{\bibfnamefont{P.~G.}
  \bibnamefont{Debenedetti}}, \bibinfo{journal}{Phys. Rev. Lett.}
  \textbf{\bibinfo{volume}{77}}, \bibinfo{pages}{4386} (\bibinfo{year}{1996}).

\bibitem[{\citenamefont{Silverstein et~al.}(1999)\citenamefont{Silverstein,
  Haymet, and Dill}}]{SilversteinHaymetDill1999}
\bibinfo{author}{\bibfnamefont{K.~A.~T.} \bibnamefont{Silverstein}},
  \bibinfo{author}{\bibfnamefont{A.~D.~J.} \bibnamefont{Haymet}},
  \bibnamefont{and} \bibinfo{author}{\bibfnamefont{K.~A.} \bibnamefont{Dill}},
  \bibinfo{journal}{J. Chem. Phys.} \textbf{\bibinfo{volume}{111}},
  \bibinfo{pages}{8000} (\bibinfo{year}{1999}).

\bibitem[{\citenamefont{Kikuchi}(1951)}]{Kikuchi1951}
\bibinfo{author}{\bibfnamefont{R.}~\bibnamefont{Kikuchi}},
  \bibinfo{journal}{Phys. Rev.} \textbf{\bibinfo{volume}{81}},
  \bibinfo{pages}{988} (\bibinfo{year}{1951}).

\bibitem[{\citenamefont{An}(1988)}]{An1988}
\bibinfo{author}{\bibfnamefont{G.}~\bibnamefont{An}}, \bibinfo{journal}{J.
  Stat. Phys.} \textbf{\bibinfo{volume}{52}}, \bibinfo{pages}{727}
  (\bibinfo{year}{1988}).

\bibitem[{\citenamefont{Morita}(1972)}]{Morita1972}
\bibinfo{author}{\bibfnamefont{T.}~\bibnamefont{Morita}}, \bibinfo{journal}{J.
  Math. Phys.} \textbf{\bibinfo{volume}{13}}, \bibinfo{pages}{115}
  (\bibinfo{year}{1972}).

\bibitem[{\citenamefont{Pretti}(2003)}]{Pretti2003}
\bibinfo{author}{\bibfnamefont{M.}~\bibnamefont{Pretti}}, \bibinfo{journal}{J.
  Stat. Phys.} \textbf{\bibinfo{volume}{11}}, \bibinfo{pages}{993}
  (\bibinfo{year}{2003}).

\bibitem[{\citenamefont{Oates et~al.}(1999)\citenamefont{Oates, Zhang, Chen,
  and Chang}}]{Oates1999}
\bibinfo{author}{\bibfnamefont{W.~A.} \bibnamefont{Oates}},
  \bibinfo{author}{\bibfnamefont{F.}~\bibnamefont{Zhang}},
  \bibinfo{author}{\bibfnamefont{S.~L.} \bibnamefont{Chen}}, \bibnamefont{and}
  \bibinfo{author}{\bibfnamefont{Y.~A.} \bibnamefont{Chang}},
  \bibinfo{journal}{Phys. Rev. B} \textbf{\bibinfo{volume}{59}},
  \bibinfo{pages}{11221} (\bibinfo{year}{1999}).

\bibitem[{\citenamefont{Kikuchi}(1974)}]{Kikuchi1974}
\bibinfo{author}{\bibfnamefont{R.}~\bibnamefont{Kikuchi}}, \bibinfo{journal}{J.
  Chem. Phys.} \textbf{\bibinfo{volume}{60}}, \bibinfo{pages}{1071}
  (\bibinfo{year}{1974}).

\bibitem[{\citenamefont{Buzano and Pretti}(2003)}]{BuzanoPretti2003jcp}
\bibinfo{author}{\bibfnamefont{C.}~\bibnamefont{Buzano}} \bibnamefont{and}
  \bibinfo{author}{\bibfnamefont{M.}~\bibnamefont{Pretti}},
  \bibinfo{journal}{J. Chem. Phys.} \textbf{\bibinfo{volume}{119}},
  \bibinfo{pages}{3791} (\bibinfo{year}{2003}).

\bibitem[{\citenamefont{F.~Sciortino}(2003)}]{SciortinoLaNaveTartaglia2003}
\bibinfo{author}{\bibfnamefont{P.~T.} \bibnamefont{F.~Sciortino},
  \bibfnamefont{E.~La~Nave}}, \bibinfo{journal}{Phys. Rev. Lett.}
  \textbf{\bibinfo{volume}{91}}, \bibinfo{pages}{155701}
  (\bibinfo{year}{2003}).

\bibitem[{\citenamefont{Speedy}(1982{\natexlab{a}})}]{Speedy1982I}
\bibinfo{author}{\bibfnamefont{R.~J.} \bibnamefont{Speedy}},
  \bibinfo{journal}{J. Phys. Chem.} \textbf{\bibinfo{volume}{86}},
  \bibinfo{pages}{982} (\bibinfo{year}{1982}{\natexlab{a}}).

\bibitem[{\citenamefont{Speedy}(1982{\natexlab{b}})}]{Speedy1982II}
\bibinfo{author}{\bibfnamefont{R.~J.} \bibnamefont{Speedy}},
  \bibinfo{journal}{J. Phys. Chem.} \textbf{\bibinfo{volume}{86}},
  \bibinfo{pages}{3002} (\bibinfo{year}{1982}{\natexlab{b}}).

\bibitem[{\citenamefont{Speedy}(1987)}]{Speedy1987}
\bibinfo{author}{\bibfnamefont{R.~J.} \bibnamefont{Speedy}},
  \bibinfo{journal}{J. Phys. Chem.} \textbf{\bibinfo{volume}{91}},
  \bibinfo{pages}{3354} (\bibinfo{year}{1987}).

\bibitem[{\citenamefont{Debenedetti and
  D'Antonio}(1986)}]{DebenedettiDantonio1986I}
\bibinfo{author}{\bibfnamefont{P.}~\bibnamefont{Debenedetti}} \bibnamefont{and}
  \bibinfo{author}{\bibfnamefont{M.~C.} \bibnamefont{D'Antonio}},
  \bibinfo{journal}{J. Chem. Phys.} \textbf{\bibinfo{volume}{84}},
  \bibinfo{pages}{3339} (\bibinfo{year}{1986}).

\bibitem[{\citenamefont{Debenedetti and
  D'Antonio}(1987)}]{DantonioDebenedetti1987}
\bibinfo{author}{\bibfnamefont{P.}~\bibnamefont{Debenedetti}} \bibnamefont{and}
  \bibinfo{author}{\bibfnamefont{M.~C.} \bibnamefont{D'Antonio}},
  \bibinfo{journal}{J. Chem. Phys.} \textbf{\bibinfo{volume}{86}},
  \bibinfo{pages}{2229} (\bibinfo{year}{1987}).

\bibitem[{\citenamefont{Imre et~al.}(2002)\citenamefont{Imre, Maris, and
  Williams}}]{Speedy2002}
\bibinfo{editor}{\bibfnamefont{A.~R.} \bibnamefont{Imre}},
  \bibinfo{editor}{\bibfnamefont{H.~J.} \bibnamefont{Maris}}, \bibnamefont{and}
  \bibinfo{editor}{\bibfnamefont{P.~R.} \bibnamefont{Williams}}, eds.,
  \emph{\bibinfo{title}{Liquids Under Negative Pressure}}
  (\bibinfo{publisher}{Kluwer Academic Publisher, Boston},
  \bibinfo{year}{2002}).

\bibitem[{\citenamefont{Zheng et~al.}(1991)\citenamefont{Zheng, Durben, Wolf,
  and Angell}}]{ZhengDurbenWolfAngell1991}
\bibinfo{author}{\bibfnamefont{Q.}~\bibnamefont{Zheng}},
  \bibinfo{author}{\bibfnamefont{D.~J.} \bibnamefont{Durben}},
  \bibinfo{author}{\bibfnamefont{G.~H.} \bibnamefont{Wolf}}, \bibnamefont{and}
  \bibinfo{author}{\bibfnamefont{C.~A.} \bibnamefont{Angell}},
  \bibinfo{journal}{Science} \textbf{\bibinfo{volume}{254}},
  \bibinfo{pages}{829} (\bibinfo{year}{1991}).

\bibitem[{\citenamefont{Truskett et~al.}(1999)\citenamefont{Truskett,
  Debenedetti, Sastry, and Torquato}}]{TruskettDebenedettiSastryTorquato1999}
\bibinfo{author}{\bibfnamefont{T.~M.} \bibnamefont{Truskett}},
  \bibinfo{author}{\bibfnamefont{P.~G.} \bibnamefont{Debenedetti}},
  \bibinfo{author}{\bibfnamefont{S.}~\bibnamefont{Sastry}}, \bibnamefont{and}
  \bibinfo{author}{\bibfnamefont{S.}~\bibnamefont{Torquato}},
  \bibinfo{journal}{J. Chem. Phys.} \textbf{\bibinfo{volume}{111}},
  \bibinfo{pages}{2647} (\bibinfo{year}{1999}).

\end{thebibliography}

\end{document}